\documentstyle[twoside,fleqn,espcrc2]{article}

\input epsf

\def\a{\alpha}
\def\b{\beta}
\def\c{\gamma}

\def\e{\epsilon}

\def\l{\lambda}
\def\m{\mu}
\def\n{\nu}

\def\r{\rho}
\def\s{\sigma}

\def\C{\Gamma}
\def\D{\Delta}
\def\L{\Lambda}

\def\pl{\partial}

\newcommand{\be}{\begin{equation}}
\newcommand{\ee}{\end{equation}}
\newcommand{\bea}{\begin{eqnarray}}
\newcommand{\eea}{\end{eqnarray}}

\newcommand{\hD}[1]{\stackrel{\leftrightarrow}{D^{#1}}}

\newcommand{\AmS}{{\protect\the\textfont2
  A\kern-.1667em\lower.5ex\hbox{M}\kern-.125emS}}

\title{The Angular Momentum and $g_1^p$ Sum Rules for the Proton\thanks
{Research supported in part by PPARC grant GR/L56374 and EC TMR grant
FMRX-CT96-0008.}  
\thanks{Invited talk at QCD00, Montpellier, July 2000.}
\thanks{CERN-TH/2000-214, SWAT/00-263, UGVA-DPT-00-7-1087.}}

\author{G.M. Shore \address{D\'epartement de Physique Th\'eorique,
Universit\'e de Gen\`eve, \\
24, quai E. Ansermet, \\
CH-1211 Geneva 4, Switzerland \\
{\rm and} \\
TH Division, CERN, \\
CH 1211 Geneva 23, Switzerland. \\    }
\thanks{Permanent address: Department of Physics, University of Wales Swansea,
Singleton Park, Swansea SA2 8PP, U.K. } 
}

\begin{document}
 
\begin{abstract}
The gauge invariant operator formulation of the angular momentum
sum rule ${1\over2} = J_q + J_g$ for the proton is presented and 
contrasted with 
the sum rule for the first moment of the polarised structure function
$g_1^p$. The decoupling of the axial charge $a^0$ from the angular momentum
sum rule is highlighted and the possible QCD field-theoretic basis for
an angular momentum sum rule of the form ${1\over2} =
{1\over2}\D q + \D g + L_q + L_g$ is critically discussed.

\end{abstract}
 
\maketitle
 
\section{Introduction}

In this talk, based on work in collaboration with B.~White, I review
our recent formulation\cite{SW} of the angular momentum sum rule for the 
proton and discuss its relation to the sum rule for the first
moment of the polarised structure function $g_1^p$. In particular, the
role of the axial charge $a^0$, which is measured in the $g_1^p$ sum rule,
is highlighted and it is shown how this decouples from the angular momentum
sum rule. This emphasises the limitations of attempting to identify $a^0$
with quark (and gluon) spin and an alternative interpretation in terms
of topological charge is briefly reviewed. 
The angular momentum sum rule is shown to take the simple form
${1\over2} = J_q + J_g$, where $J_q$ and $J_g$ are gauge and Lorentz
invariant form factors of the forward matrix elements of local operators,
which may reasonably be interpreted as quark and gluon components
of the total angular momentum of the proton. Experimentally, they may be 
measured in, for example, deeply virtual Compton scattering. 
Their RG evolution properties are derived from a careful analysis of
operator mixing. 
We also discuss critically whether
there is indeed a QCD field-theoretic basis for a decomposition of the
proton spin into separate quark and gluon spin and orbital angular momentum
components, as in the frequently-quoted sum rule ${1\over2} =
{1\over2}\D q + \D g + L_q + L_g$.

\section{The $g_1^p$ sum rule and topological charge}

The sum rule for the first moment of $g_1^p$ (see e.g.~ref.\cite{Srev} for 
reviews of our earlier work and references) is
\be
\int_0^1 dx~ g_1^p(x,Q^2) = {1\over12}C_1^{\rm NS} \bigl(a^3 + {1\over3}a^8
\bigr) + {1\over9} C_1^{\rm S} a^0(Q^2)~~~
\ee
where the $C_1$ are Wilson coefficients and the flavour singlet axial 
charge $a^0$ is defined as the form factor in the forward matrix element
of the corresponding axial current,
\be
\langle p,s|A_\mu^0|p,s\rangle = a^0 s_\mu
\ee
where $s_\mu$ is the covariant spin vector.
Since $A_\mu^0$ is not a conserved current, due to the $U_A(1)$ anomaly
\be
\pl^\mu A_\mu^0 - 2n_f Q \sim 0
\ee
where $Q = {\a_s\over8\pi}\e^{\m\n\r\s}{\rm tr} F_{\m\n} F_{\r\s}$
is the gluon topological charge density, the form factor $a^0$ is scale
dependent and satisfies the non-trivial RG evolution equation
\be
{d\over dt} a^0(Q^2) = \c a^0(Q^2)
\ee
where $t= \ln Q^2/\L^2$ and the anomalous dimension is 
$\c = -n_f {\a_s\over2\pi^2}$.

Given $a^3$ and $a^8$ from low-energy neutron and hyperon decays,
$a^0$ can be extracted from polarised inclusive DIS processes
$e(\m)p \rightarrow e(\m) X$. The simplest prediction, $a^0 \simeq a^8$,
is an immediate consequence of the OZI rule and transcribed into the
sum rule for $g_1^p$ gives the Ellis-Jaffe sum rule. 
However, the flavour singlet pseudovector or pseudoscalar channel is
precisely where we would expect to find strong OZI violations related to
the $U_A(1)$ anomaly and indeed it is found experimentally that
$a^0 \ll a^8$.

Since we can rewrite $a^0$ using eq.(3) in terms of the matrix element
of the topological charge density,
\be
a^0 = \langle p,s| Q | p,s\rangle
\ee
we see immediately that the observed suppression in $a^0$ is a manifestation
of topological charge screening. This interpretation has been developed in
a series of papers written in collaboration with Veneziano, Narison and 
De Florian\cite{Srev}. Our proposal is that this screening is 
{\it universal}, i.e.~target-independent, being an intrinsic property of the
QCD vacuum itself. Specifically, we showed that (in the chiral limit)
\be
a^0 = {1\over2M} 2n_f\Bigl[ \chi(0)\C_{Qpp} + \sqrt{\chi^\prime(0)}
\C_{\Phi_5 pp} \Bigr]
\ee
where the $\C$ are suitably defined 1PI vertex functions and
$\chi(k^2) = i \int d^4x e^{ik.x} \langle 0|T Q(x) Q(0)|\rangle$
is the {\it topological susceptibility}, $\chi^\prime(0)$ being its slope
at $k=0$. Since the anomalous chiral Ward identity implies $\chi(0)=0$
in the chiral limit, the sole contribution to $a^0$ comes from the second
term in eq.(6). Making the motivated assumption that the RG invariant vertex
$\C_{\Phi_5 pp}$ obeys the OZI rule to a good approximation, we conjecture
that the principal origin of the suppression in $a^0$ is an anomalously
small value of $\chi^\prime(0)$ due to universal topological charge screening
by the QCD vacuum.\footnote{
Explanations which favour OZI violations due to a large
polarised strange quark component of the proton or the implications
of the Skyrme model of proton structure have been recently reviewed
in, for example, ref.\cite{Ellis} }  This is anticipated in certain 
instanton-based models of the QCD vacuum, has been confirmed quantitatively by 
QCD spectral sum rule calculations,
and is currently being investigated in lattice gauge theory.

The QCD parton model gives an alternative interpretation of $a^0$ through
the identification 
\be
a^0(Q^2) = \D q - 2n_f {\a_s\over4\pi} \D g(Q^2)
\ee
where $\D q = \D u + \D d + \D s$ and $\D g(Q^2)$ are the first moments
of the polarised flavour singlet quark and gluon distributions
and we have used the AB class of renormalisation schemes where 
$\D q$ is defined to be $Q^2$ independent. 
The OZI/Ellis-Jaffe relation $a^0=a^8$ follows immediately from the
assumption that in the proton, $\D s = 0$ and $\D g = 0$.
In this model, ${1\over2} \D q$ and $\D g$ are interpreted as the quark 
and gluon spins, which led to the initial interpretation of the
experimental observation $a^0 \ll a^8$ as indicating that the quarks carry 
only a small fraction of the spin of the proton -- the so-called `proton
spin crisis'.

In the rest of this talk, we derive the actual angular momentum sum rule 
for the proton in terms of gauge invariant operator matrix elements,
with particular emphasis on whether and how the axial charge $a^0$
appears and whether the interpretation of $\D q$ and $\D g$ as spin 
components can be complemented by corresponding definitions of orbital
angular momentum components $L_q$ and $L_g$.

\section{The angular momentum sum rule}

The angular momentum sum rule is derived by taking the forward matrix element
of the conserved angular momentum current $M^{\m\n\l}$, defined from
the energy-momentum tensor as
\be
M^{\m\n\l} = x^{[\n} T^{\l]\m} + \partial_\r X^{\r\m\n\l}
\ee
The inclusion of the arbitrary tensor $X^{\r\m\n\l}$ (antisymmetric under 
$\r\leftrightarrow\m$ and $\n\leftrightarrow\l$) just reflects the 
usual freedom in QFT in defining conserved currents. However, this 
arbitrariness allows us to write different equivalent expressions for
$M^{\m\n\l}$ as a sum of local operators, suggesting corresponding 
interpretations of the total angular momentum as a sum of `components',
which we can try to identify as reasonable definitions of quark and gluon
spin and orbital angular momentum\cite{Jaffe}. 

The best decomposition is the following:
\bea
M^{\m\n\l} = O_1^{\m\n\l} + O_2^{\m[\l}x^{\n]} + O_3^{\m[\l}x^{\n]}
+ g^{\m[\l}x^{\n]} {\cal L}_{\rm gi} \nonumber \\
+ \big\{i\pl^{\{\m}\bar c D^{[\l\}}c + \pl^{\{\m}B A^{[\l\}}
+ g^{\m[\l} {\cal L}_{\rm gf} \big\}x^{\n]} \nonumber \\
-{1\over4}\pl_\r\big[x^{[\n} \e^{\l]\m\r\s}\bar\psi \c_\s\c_5\psi\big\}
+ {\rm EOM} + \partial_\r X^{\r\m\n\l}   \\
\nonumber
\eea
The tensor $X^{\r\m\n\l}$ is chosen to cancel the divergence and 
equation of motion (EOM) terms, while the forward matrix element of 
the operator $g^{\m[\l}x^{\n]} {\cal L}_{\rm gi}$ vanishes. The term in
$\{ \ldots \}$ is the contribution from the covariant gauge-fixing and ghost
terms in the Lagrangian, but this turns out to be a BRS variation so
its matrix element between physical states vanishes. The remaining (bare)
operators are gauge invariant:
\bea
&{}&O_1^{\m\n\l} = {1\over2} \e^{\m\n\l\s}\bar\psi \c_\s \c_5\psi
\equiv {1\over2} \e^{\m\n\l\s} A_\s^0 \nonumber \\
&{}&O_2^{\m\l} = i \bar\psi \c^\m \hD{\l} \psi \nonumber \\
&{}&O_3^{\m\l} = F^{\m\r} F_{\r}{}^{\l}    \\
\nonumber 
\eea
We also find it convenient for later use to define $O_4^{\m\l} =
{1\over2} O_2^{\{\m\l\}}$.
At first sight, $O_1^{\m\n\l}$, which is just the flavour singlet axial
current considered in section 2, looks as if it may be associated with 
`quark spin', with $O_2^{\m[\l}x^{\n]}$ corresponding to a gauge invariant
definition of `quark orbital angular momentum'. This leaves 
$O_3^{\m[\l}x^{\n]}$ to be associated with the `gluon total angular momentum'.

In fact, there is no further decomposition of the gluon contribution as
long as we restrict to gauge invariant operators. We can certainly make 
the alternative decomposition:
\bea
M^{\m\n\l} = ~~~~~~~~~~~~~~~~~~~~~~~~~~~~~~~~~~~~~~~~~~~~~~~~~~~~~\nonumber \\
\tilde O_1^{\m\n\l} + \tilde O_2^{\m[\l}x^{\n]} 
+ \tilde O_3^{\m\n\l} + \tilde O_4^{\m[\l}x^{\n]}
+ g^{\m[\l}x^{\n]} {\cal L}_{\rm gi} \nonumber \\
+ \big\{A^\m \pl^{[\l}B + i \pl^\m \bar c \pl^{[\l} c 
+ i \pl^{[\l}\bar c D^\m c
+ g^{\m[\l} {\cal L}_{\rm gf} \big\}x^{\n]} \nonumber \\
+\pl_\r\big[x^{[\n} A^{\l]} F^{\m\r}\bigr]
+ {\rm EOM} + \partial_\r X^{\r\m\n\l}~~~~~~~~~~~~~~~~   \\ 
\nonumber
\eea
where
\bea 
&{}&\tilde O_1^{\m\n\l} = {1\over2} \e^{\m\n\l\s}\bar\psi \c_\s\c_5\psi 
\nonumber \\
&{}&\tilde O_2^{\m\l} = -i \bar\psi \c^\m \pl^\l \psi \nonumber \\
&{}&\tilde O_3^{\m\n\l} = -F^{\m[\n}A^{\l]} \nonumber \\
&{}&\tilde O_4^{\m\n} = F^{\m\r} \pl^\l A_\r   \\
\nonumber
\eea
and try to identify the first four operators with, respectively, 
quark spin and orbital and gluon spin and orbital angular momentum.
However, even the forward matrix elements of these operators turn out 
not to be gauge invariant. Moreover, the gauge-fixing and ghost term
in $\{\ldots\}$ is no longer a BRS variation so would contribute
a non-vanishing `ghost orbital angular momentum'. Further discussion
of the problems with such gauge non-invariant decompositions is given
in ref.\cite{SW}, and from now on we restrict attention to the gauge
invariant formulation based on eq.(9).

The next step is to express the matrix elements of the operators 
$O_1^{\m\n\l}$, $O_2^{\m[\l}x^{\n]}$ and $O_3^{\m[\l}x^{\n]}$ in terms
of form factors. There are technical subtleties connected with defining
operators of the form $Ox$ and their renormalisation mixing which are
explained precisely in ref.\cite{SW}. The prescription is essentially to 
define the forward matrix elements of $Ox$ in terms of the limit of an 
off-forward matrix element. We therefore write (a little loosely)
\be
\langle p|O^{\m\l}x^\n|p\rangle = -i{\pl\over\pl\D_\n}
\langle p|O^{\m\l}|p'\rangle\big|_{p'=p}
\ee
where $\D = p-p'$. We then have\cite{SW}
\bea
\langle p,s|O_1^{\m\n\l}|p,s\rangle = 
a^0 M\e^{\m\n\l\s}s_\s~~~~~~~~~~~~~~~~~~~~ 
\nonumber \\
\langle p,s|O_2^{\m[\l}x^{\n]}|p,s\rangle = 
B_q(0) {1\over2M} p_\r p^{\{\m} \e^{[\l\}\n]\r\s}s_\s\nonumber \\
+\tilde B_q(0) {1\over2M} p_\r p^{[\m} \e^{[\l]\n]\r\s}s_\s 
- 2D_q(0) M\e^{\m\n\l\s}s_\s \nonumber \\
\langle p,s|O_3^{\m[\l}x^{\n]}|p,s\rangle = 
B_g(0) {1\over2M} p_\r p^{\{\m} \e^{[\l\}\n]\r\s}s_\s   \\
\nonumber
\eea
The crucial observation now follows from the identity
\be
O_1^{\m\n\l} + O_2^{\m[\l}x^{\n]} =
O_4^{\m[\l}x^{\n]} + {\rm divergence} + {\rm EOM}~~~
\ee
Since the matrix elements of the divergence and {\rm EOM} terms vanish,
and recalling that the operator $O_4^{\m\l}$ is symmetric in $\m,\l$,
we see that
\be 
\tilde B_q(0) = 0 ~~~~~~~~~~~~~~~~~2D_q(0) = a^0
\ee
Thus $a^0$ appears not only as the unique form factor in the matrix element
$\langle O_1\rangle$ of the axial current, but also as a contribution
to $\langle O_2 x\rangle$. It therefore {\it cancels} from the angular momentum
sum rule.

Introducing the notation $J_q = B_q(0)$, $J_g = B_g(0)$, then from eq.(14)
we may write the sum rule as
\be
{1\over2} = J_q + J_g
\ee
where $J_q$ and $J_g$ are gauge and Lorentz invariant quantities which
may reasonably be identified as total `quark' and `gluon' angular momenta
respectively. Of course, this is a rather non-rigorous terminology
since the corresponding operators $O_2x$ and $O_3x$ mix, and indeed $O_2x$
itself contains an explicit gluon field component in the covariant derivative,
but it is convenient and the closest approximation to a quark-gluon
decomposition that can be given in an interacting QFT such as QCD.
Moreover, as we discuss below, $J_q$ and $J_g$ are still not RG scheme/scale
dependent quantities.

The point we wish to stress is that provided we restrict to gauge and
Lorentz invariant quantities, the true angular momentum sum rule
involves only the two form factors $J_q$ and $J_g$. The axial charge $a^0$ is
simply not present in the sum rule (17). We return to this point in section 5.

Just as the axial charge form factor $a^0$ can be measured in polarised 
inclusive DIS, the angular momentum form factors $J_q$ and $J_g$ can in
principle be extracted from measurements of unpolarised off-forward parton
distribution functions in processes such as deeply virtual Compton scattering
$\c^* p \rightarrow \c p$. The required identifications are
\bea
-iP^+ {\pl\over\pl \D_\m} \int_{-1}^{1} dx~ x f_{q(g)/p}(x,\xi,\D)\bigl|_{\D=0}
\nonumber \\
= J_{q(g)} {1\over M} \e^{+\m\r\s} P_\r s_\s  \\
\nonumber
\eea
where $\xi = {q.\D\over2q.P}$ and the incoming(outgoing) proton momenta
are $P-(+)\D$.

These form factors may also be calculated non-perturbatively in lattice 
gauge theory and some initial results for $J_q$ in the quenched approximation
have recently been obtained in ref.\cite{Dong}.

\section{Operator mixing and RG evolution}

The operators $O_1$, $O_2x$ and $O_3x$ in the angular momentum sum rule 
renormalise and mix in a non-trivial way. The analysis is made more subtle
by the explicit factors of the coordinate $x$ which have to be carefully
treated. A detailed discussion is presented in ref.\cite{SW} and here
we only sketch the main features.

First, note that when inserted into forward matrix elements, operators of the
form $O_a$ and $O_ix$ mix with a block triangular structure:
\be
  \left( \begin{array}{c} O_a \\ O_ix \end{array} \right)_R
   = \left( \begin{array}{cc}  Z_{ab}^{-1} & 0 \\ Z_{ib}^{-1} & Z_{ij}^{-1}
   \end{array} \right)
   \left( \begin{array}{c} O_b \\ O_jx \end{array} \right)_B
\ee
since gauge-invariant operators with no factors of $x$ only mix with other
similar operators. Then since $O_3^{\m\l}$ is symmetric, it can only mix with 
the symmetric operators $O_3^{\m\l}$ and $O_4^{\m\l}$, which for forward matrix
elements implies that $O_3^{\m[\l}x^{\n]}$ only mixes with itself
and $O_1^{\m\n\l} + O_2^{\m[\l}x^{\n]}$. Finally, since the full angular momentum
current is conserved and therefore not renormalised, the columns of the mixing
matrix must all add to one. This implies the following form for the mixing matrix
for forward matrix elements:
\bea
  \left( \begin{array}{c} O_1^{\mu\nu\lambda} \\ O_2^{\mu[\lambda} x^{\nu]}
   \\ O_3^{\mu[\lambda} x^{\nu]} 
   \end{array}\right)_B = ~~~~~~~~~~~~~~~~~~~~~~~~~~~~~
   \nonumber \\
   \left(\begin{array}{ccc} 1 + X & 0 & 0 \\
                            Z - X & 1 + Z & -Y \\
                            -Z & -Z & 1 + Y \end{array}\right)
   \left(\begin{array}{c} O_1^{\mu\nu\lambda} \\ O_2^{\mu[\lambda} x^{\nu]}
   \\ O_3^{\mu[\lambda} x^{\nu]} \end{array}\right)_R  ~~  \\
\nonumber
\eea
and one-loop calculations show $Y= -{2\over3}n_f {\a_s\over4\pi}{1\over\e}$
and $Z= -{8\over3}C_F {\a_s\over4\pi}{1\over\e}$. $X$ is due to the anomaly and
is $O(\a_s^2)$. For the form factors, this gives:
\bea
  \left( \begin{array}{c} a^0 \\ B_q
   \\ B_g \end{array}\right)_B =  ~~~~~~~~~~~~~~~~~~~~~~~~~~~~~~~
   \nonumber \\
   \left(\begin{array}{ccc} 1 + X & 0 & 0 \\
                            0 & 1 + Z & -Y \\
                            0 & -Z & 1 + Y \end{array}\right)
   \left(\begin{array}{c} a^0 \\ B_q
   \\ B_g \end{array}\right)_R  ~~ \\
\nonumber   
\eea

We therefore find the evolution equations for the quark and gluon
components of the proton angular momentum:
\be
{d\over dt} 
\left( \begin{array}{c} J_q \\ J_g \end{array}\right) ={\a_s\over4\pi}
   \left(\begin{array}{cc} -{8\over3}C_F& {2\over3}n_f \\
                            {8\over3}C_F& -{2\over3}n_f
                            \end{array}\right)
   \left(\begin{array}{c} J_q \\ J_g \end{array}\right)~~~
\ee
together with eq.(4) for $a^0$. It follows that 
in the asymptotic limit $Q^2\rightarrow \infty$, the partitioning of quark:gluon
angular momenta is $3n_f:16$ \cite{HJT}. Interestingly, this is the same
result as for the partitioning of momentum obtained from the first moment
of the unpolarised pdfs.

\section{Partons and orbital angular momentum}

We have seen how the $g_1^p$ and angular momentum sum rules involve three
Lorentz invariant form factors $a^0$, $J_q$ and $J_g$, where 
$J_{q(g)}$ may be reasonably identified as total quark (gluon) 
angular momentum components in the sum rule ${1\over2} = J_q + J_g$
while the axial charge $a^0$, which enters the $g_1^p$ sum rule, decouples
and may instead be interpreted in terms of topological charge density.
There is no gauge and Lorentz invariant operator identification of
`orbital angular momenta' $L_q$ and $L_g$.

In the parton model, the axial charge $a^0$ is interpreted as a sum of
polarised quark $\D q$ and gluon $\D g$ distributions as in eq.(7).
In the AB scheme, the RG evolution equations are
\be
{d\D q\over dt} = 0 ~~~~~~~~~~~~~~~{d\D g\over dt} = {\a_s\over4\pi}
\bigl(3C_F \D q + \b_0 \D g\bigr)~~~
\ee
where $\b_0 = 11 - {2\over3} n_f$, compatible with eq.(4).
This evolution can also be directly obtained from the splitting functions.

Now if as the parton model suggests, ${1\over2} \D q$ and $\D g$ are to be 
interpreted as quark and gluon spins, then to complete the angular 
momentum sum rule we are forced to write
\be
{1\over2} = {1\over2} \D q + \D g + L_q + L_g
\ee
where $L_{q(g)}$ are orbital angular momenta. Since there is
no intrinsic operator definition of these partonic quantities, 
the best we can do is
to define them to be consistent with the sum rule (17), i.e.
\be
L_q = J_q - {1\over2}\D q  ~~~~~~~~~~
L_g = J_g - \D g 
\ee
Of course this means the sum rule (24) is not really predictive, since there is 
no way to independently measure $L_{q(g)}$ -- only the form factors
$J_{q(g)}$ can be extracted from experiment. Moreover, the identifications
(25) are not very natural, since they involve subtracting quantities belonging to
form factors for different Lorentz structures. They are therefore
frame-dependent, not surprisingly given that spin and
orbital angular momentum are associated with different representations
of the Lorentz group.

Nevertheless, if we adopt (7),(25),(24) as the best possible 
gauge-invariant definition of a quark/gluon spin/orbital angular momentum 
sum rule, we may determine the RG evolution for the components
$\D q$, $\D g$, $L_q$ and $L_g$ from eqs. (22),(23).  
We find
\bea
  \frac{d}{dt} \left(\begin{array}{c} \Delta q \\ \Delta g \\ L_q \\ L_g
   \end{array}\right) =   ~~~~~~~~~~~~~~~~~~~~~~~~~~~~~~~~~~~~~~~~~~~~
    \nonumber \\
   \frac{\alpha_s}{4\pi} \left(
   \begin{array}{cccc} 0 & 0 & 0 & 0 \\
                3C_F & \beta_0 & 0 & 0 \\
          -\frac{4}{3}C_F & \frac{2}{3}n_f & -\frac{8}{3}C_F & \frac{2}{3}n_f\\
          -\frac{5}{3}C_F & -11 & \frac{8}{3}C_F & -\frac{2}{3}n_f
   \end{array}\right)\left(\begin{array}{c}\Delta q \\ \Delta g \\ L_q \\ L_g
   \end{array}\right) \\
\nonumber   
\eea
which is consistent with the non-renormalisation of the
full angular momentum current and may, at least in part, also be derived
in a splitting function approach \cite{HJT}.

\end{document}